\documentclass[12pt,english,floatfix,showkeys,superscriptaddress,aps,prd,preprint]{revtex4}
\usepackage[T1]{fontenc}
\usepackage{lmodern}
\usepackage{amsmath}
\usepackage{amssymb}
\usepackage{graphicx}
\usepackage{float}
\usepackage{esint}
\usepackage{caption}
\usepackage{subcaption}
\usepackage{dcolumn}
\usepackage{babel}
\usepackage{csquotes}
\usepackage{color}
\usepackage{slashed}
\usepackage{simplewick}
\usepackage{tikz-feynman}
\usepackage[]{tikz-feynman}
\usepackage{amsmath,latexsym}

\usepackage{hyperref}
\hypersetup{
    colorlinks,
    citecolor=blue,
    filecolor=green,
    linkcolor=purple,
    urlcolor=red,
}

\usepackage{slashed}

\usepackage{hyperref}
\hypersetup{colorlinks,breaklinks,
			citecolor=[rgb]{0,0.0,1.0},
            urlcolor=[rgb]{0.0,0.0,1.0},
            linkcolor=[rgb]{0,0.5,0.9}}

\begin{document}

\title{Perturbative solutions for compact objects in (2+1)-dimensional Bopp-Podolsky electrodynamics}

\author{R. V. Maluf}
\email{r.v.maluf@fisica.ufc.br}
\affiliation{Universidade Federal do Cear\'a (UFC), Departamento de F\'isica,\\ Campus do Pici, Fortaleza - CE, C.P. 6030, 60455-760 - Brazil.}
\author{J. E. G. Silva}
\email{euclides@fisica.ufc.br}
\affiliation{Universidade Federal do Cear\'a (UFC), Departamento de F\'isica,\\ Campus do Pici, Fortaleza - CE, C.P. 6030, 60455-760 - Brazil.}
\author{C. A. S. Almeida}
\email{carlos@fisica.ufc.br}
\affiliation{Universidade Federal do Cear\'a (UFC), Departamento de F\'isica,\\ Campus do Pici, Fortaleza - CE, C.P. 6030, 60455-760 - Brazil.}

\author{Gonzalo J. Olmo}
\email{gonzalo.olmo@uv.es}
\affiliation{Departamento de F\'{i}sica Te\'{o}rica and IFIC, Centro Mixto Universidad de Valencia - CSIC. Universidad
de Valencia, Burjassot-46100, Valencia, Spain.}
\affiliation{Universidade Federal do Cear\'a (UFC), Departamento de F\'isica,\\ Campus do Pici, Fortaleza - CE, C.P. 6030, 60455-760 - Brazil.}


\date{\today}

\begin{abstract}
We investigate the space-time geometry generated by compact objects in (2+1)-dimensional Bopp-Podolsky electrodynamics. Inspired by previous studies where the Bopp-Podolsky field acts as a source for spherically symmetric solutions, we revisit this question within the lower-dimensional (2+1) framework. Using a perturbative approach, we derive a charged BTZ-like black hole solution and compute corrections up to second order in a perturbative expansion valid far from the horizon. Our analysis suggests that the near-horizon and inner structure of the solution remain unaltered, indicating that no new non-black hole objects emerge in this regime. In particular, we do not find evidence of wormhole solutions in the (2+1)-dimensional version of this theory.
\end{abstract}


\maketitle

\section{Introduction}\label{intro}

The Bopp--Podolsky electrodynamics \cite{Bopp1940,Podolsky:1942zz,Podolsky:1944zz} extends Maxwell’s theory by incorporating second-order derivatives of the gauge field into the Lagrangian. This modification introduces higher-order field equations, characterized by fourth-order derivative terms. A notable feature of this theory is the inclusion of a massive mode while preserving gauge symmetry, which sets it apart from the standard Proca model \cite{Proca,Tu:2005ge}. Nevertheless, it is well known that higher-derivative models often exhibit dynamic instabilities. At the classical level, the Noether energy is typically unbounded from below, and when quantum properties are taken into account, higher-derivative terms lead to the emergence of ghost instabilities \cite{Accioly:2010zza}. However, it can be shown that such theories possess a positive (non-canonical) energy by employing the concept of a Lagrange anchor (for further details, see Ref. \cite{Kaparulin:2014vpa}). 

In Bopp--Podolsky electrodynamics, the standard Lorenz gauge condition is replaced by a modified gauge condition \cite{Galvao:1986yq}, which is better suited to account for the five degrees of freedom in the spectrum. Specifically, two correspond to the massless photon mode, while the remaining three are associated with the massive longitudinal mode \cite{Ferreira:2019lpu}. Studies on the Bopp--Podolsky electrodynamics have been carried out in several contexts, including radiative corrections in the low-energy regime \cite{Borges:2019gpz}, renormalization \cite{Bufalo:2012tt}, path integral quantization\cite{Bufalo:2010sb}, and finite-temperature approaches \cite{Bonin:2009je,AraujoFilho:2020bzd}, multipole expansions \cite{Bonin:2016gav}, black hole and wormhole solutions \cite{Cuzinatto:2017srn,Frizo:2022jyz}, cosmology \cite{Cuzinatto:2016kjk}, and several other applications \cite{Kruglov:2009yr,Cuzinatto:2011zz,Zayats:2013ioa,Granado:2019bqk}.

Recently, an interplay between the no-hair conjecture and the Bopp--Podolsky electrodynamics in curved spacetime scenarios has garnered significant attention. The no-hair conjecture, or theorem \cite{Israel:1967wq,Israel:1967za,Carter:1971zc}, asserts that any stationary black hole solution to the Einstein-Maxwell equations in general relativity can be completely characterized by only three parameters: mass, charge, and angular momentum. Cuzinatto et al. \cite{Cuzinatto:2017srn} argue that, in spherically symmetric spacetimes, only Maxwell modes propagate outside the event horizon, thereby satisfying the no-hair theorem in the context of Bopp--Podolsky fields. However, they also suggest that solutions featuring hair are not mathematically excluded when the Bopp--Podolsky parameter $b$ (associated with the gauge field mass in flat spacetime) is nonzero. 

On the other hand, Frizo et al. \cite{Frizo:2022jyz} obtained an analytic solution to the Einstein-Bopp-Podolsky gravity equations within a perturbative approach, asserting that, in the first-order approximation, the solution represents an explicit wormhole metric. Furthermore, the authors interpret the parameter $b$ as a constant of nature, rather than a property of a body, such as mass or charge. Consequently, they argue that the presence of $b$ does not contradict the no-hair conjecture.

In this context, inspired by the aforementioned works in which Bopp--Podolsky matter serves as a source for spherically symmetric solutions, we revisit this question in the framework of lower-dimensional $2+1$ spacetime. Using a perturbative approach similar to that employed in Ref. \cite{Frizo:2022jyz}, we show that a charged BTZ-like black hole solution can be derived. We compute up to second-order corrections in a perturbative expansion far from the horizon, including terms that involve the $b$ parameter. From our analysis, nothing suggests that the near-horizon or inner structure of the solution gets modified to produce non-black hole objects. Thus, we do not see any evidence regarding the possible existence of wormholes in the $2+1$ dimensional version of this theory. 

Throughout the paper, we adopt geometrized units, setting $G = c = 1$, where $G$ is the Newtonian gravitational constant in two spatial dimensions and $c$ is the speed of light in vacuum. Furthermore, the metric signature used is $(1, -1, -1)$.

\section{The Bopp--Podolsky electrodynamics in the gravitational scenario\label{S2}}

This section presents the action and field equations for the Bopp--Podolsky electrodynamics in curved spacetime. Following Cuzinatto et al. \cite{Cuzinatto:2017srn} and Frizo et al. \cite{Frizo:2022jyz}, the Lagrangian for this framework includes two additional independent and invariant terms beyond the standard Maxwell term. These contributions exhibit both minimal and non-minimal couplings to gravity. The Bopp–Podolsky electrodynamics in curved spacetime is thus governed by the following Lagrangian density:
\begin{equation}
\mathcal{L}_{m} =- \frac{1}{4}F^{\alpha\beta}F_{\alpha\beta}+\frac{\left(
a^{2}+2b^{2}\right)  }{2}\nabla_{\beta}F^{\alpha\beta}\nabla_{\gamma}%
F_{\alpha}^{\text{ \ }\gamma}
+b^{2}\left(  R_{\sigma\beta}F_{\text{ \ }}^{\sigma\alpha}F_{\alpha}^{\text{
\ }\beta}+R_{\alpha\sigma\beta\gamma}F^{\sigma\gamma}F^{\alpha\beta}\right),
\label{Lagrangian}
\end{equation}
where $F_{\mu\nu}=\nabla_{\mu}A_{\nu}-\nabla_{\nu}A_{\mu}$ is the field strength, $\nabla_{\mu}$ is
the covariant derivative,  $a$ and $b$ are coupling constants, $R_{\mu\nu}$ is the Ricci tensor, and
$R_{\alpha\beta\gamma \delta}$ is the Riemann tensor.

As we can see, the Lagrangian (\ref{Lagrangian}) is invariant under local Lorentz transformations and the $U(1)$ gauge symmetry group. Additionally, it exhibits quadratic dependence on the gauge field and its derivatives up to the fourth order.
 
We now utilize the Bopp--Podolsky electrodynamics to express the Einstein--Hilbert action for three-dimensional gravity as:
\begin{equation}
S=\frac{1}{16\pi}\int d^3x \sqrt{-g}\left(-R-2\Lambda+4 \mathcal{L}_m \right),
\label{Action}
\end{equation}
where $g$ is the metric determinant, $R$ is the Ricci scalar, $\Lambda$ is the cosmological constant, and $\mathcal{L}_{m}$ is given by Eq. (\ref{Lagrangian}).

The gravitational field equations are obtained by varying the action (\ref{Action}) with respect to the metric tensor, yielding,
\begin{equation}
R_{\mu\nu}-\frac{1}{2}g_{\mu\nu}R-\Lambda g_{\mu\nu}= 8\pi\left(  T_{\mu\nu}^{M}+T_{\mu\nu}^{a}+T_{\mu\nu}^{b}\right),
\label{Eq_gravitacional}%
\end{equation}
with the components of the energy-momentum tensor given by
\begin{eqnarray}
T_{\mu\nu}^{M} &=&\frac{1}{4\pi}\left[  F_{\mu\sigma}F_{\text{ \ }\nu
}^{\sigma}+g_{\mu\nu}\frac{1}{4}F^{\alpha\beta}F_{\alpha\beta}\right],
\label{T_munu M}\\
T_{\mu\nu}^{a} &=&\frac{a^{2}}{4\pi}\left[  g_{\mu\nu}F_{\beta}^{\;\gamma
}\nabla_{\gamma}K^{\beta}+\frac{g_{\mu\nu}}{2}K^{\beta}K_{\beta}\right.
 +\left.  2F_{(\mu}^{~\alpha}\nabla_{\nu)}K_{\alpha}-2F_{(\mu}^{~\alpha
}\nabla_{\alpha}K_{\nu)}-K_{\mu}K_{\nu}\right], \label{T_munu a}\\
T_{\mu\nu}^{b} &=&\frac{b^{2}}{2\pi}\left[  -\frac{1}{4}g_{\mu\nu}%
\nabla^{\beta}F^{\alpha\gamma}\nabla_{\beta}F_{\alpha\gamma}+F_{\text{ }(\mu
}^{\gamma}\nabla^{\beta}\nabla_{\beta}F_{\nu)\gamma}\right. 
+ \left.  F_{\gamma(\mu}\nabla_{\beta}\nabla_{\nu)}F^{\beta\gamma}%
-\nabla_{\beta}\left(  F_{\gamma}^{\text{ \ }\beta}\nabla_{(\mu}F_{\nu
)}^{\text{ \ }\gamma}\right)  \right].  \label{T_munu b}%
\end{eqnarray}
The notation $\left(  ...\right)  $ indicates symmetrization with respect to the indices inside the brackets.

Now, the Bopp--Podolsky field equations are determined by varying the action (\ref{Action}) with respect to the potential $A_{\mu}$, which yields  
\begin{equation}
\nabla_{\nu}\left[  F^{\mu\nu}-\left(  a^{2}+2b^{2}\right)  H^{\mu\nu}%
+2b^{2}S^{\mu\nu}\right]  =0,\label{Eq_Podolsky}
\end{equation}
where \begin{eqnarray}
H^{\mu\nu} &\equiv &\nabla^{\mu}K^{\nu}-\nabla^{\nu}K^{\mu},\label{H_munu}\\
S^{\mu\nu} &\equiv &F_{\text{ \ }}^{\mu\sigma}R_{\sigma}^{\text{ \ }\nu
}-F^{\nu\sigma}R_{\sigma}^{\text{ \ }\mu}+2R_{\text{ \ }\sigma\text{ }\beta
}^{\mu\text{ \ }\nu}F^{\beta\sigma},\label{S_munu}%
\end{eqnarray}
with $K^{\mu}\equiv\nabla_{\gamma}F^{\mu\gamma}$. It is worth noting that the total energy--momentum tensor also satisfies the conservation equation, i.e.,
given $T_{\mu\nu}=T_{\mu\nu}^M+T_{\mu\nu}^a+T_{\mu\nu}^b$, one has
\begin{equation}
\nabla_{\nu}T^{\mu\nu}=0.
\label{Conservation}
\end{equation} 

In the four-dimensional scenario, the search for solutions to the Einstein--Bopp--Podolsky equations was carried out in Ref. \cite{Cuzinatto:2017srn} using the Bekenstein method. However, this was done without specifying an explicit form for the spacetime metric. It was shown that for the case $b=0$, the exterior solutions to the spherically symmetric metric reproduce the Reissner--Nordstr\"{o}m solution, meaning the no-hair theorem is not violated. Furthermore, they also argue that a solution with hair
could be achieved when $b\neq 0$. On the other hand, in Ref. \cite{Frizo:2022jyz}, the authors employed a perturbative approach and have shown that only $b\neq 0$ contributes to the solution at first-order perturbation and claim that the solution obtained describes a wormhole instead of a black hole geometry. However, the authors argue that $b$ is not a black hole parameter, such as mass, charge, or spin, but rather a constant of nature, aligning its presence with the no-hair conjecture.

In the next section, we apply the same perturbative approach, extended to analyze second-order perturbations around the BTZ solution. We show that the resulting perturbative metric depends solely on the $b$-parameter, similar to its four-dimensional counterpart. However, as will be shown, this solution represents a genuine black hole geometry---specifically, a charged BTZ black hole with corrections that decay sufficiently fast in the far region but increase the curvature closer to the origin.

\section{Perturbative solutions of the Einstein--Bopp--Podolsky gravity\label{S3}}

To find solutions to the Einstein--Bopp--Podolsky field equations in $2+1$ dimensions, let us assume a static, circularly symmetric spacetime whose metric \textit{ansatz} is given by
\begin{equation}
ds^2 = A(r)dt^2-\frac{dr^2}{B(r)}-r^2 d\theta^2,
\label{SS}
\end{equation} where $A(r)$ and $B(r)$ are arbitrary functions of the radial coordinate $r$. Furthermore, since we are investigating small deviations from the charged BTZ solution, we disregard the influence of the magnetic field, considering only the electric field as the source of the gravitational field in the geometry defined by the metric \textit{ansatz} (\ref{SS}). Thus, we assume that the electromagnetic field tensor can be expressed in the simplified form:
\begin{equation}
F_{\mu\nu}=E(r) \left[ \delta^{1}_{\mu} \delta^{0}_{\nu}- \delta^{0}_{\mu} \delta^{1}_{\nu} \right],
\label{F}
\end{equation} where $E(r)$ represents the two-dimensional spatial electric field.

Unfortunately, even in $2+1$ dimensions, the nonlinearity and the inclusion of higher-derivative terms in the field equations make it impossible to obtain an exact analytical solution. As an alternative to numerical methods, a perturbative approach proves to be a viable option.

The Einstein–Bopp–Podolsky equations can be solved perturbatively around a background spacetime, assuming the perturbations are small. The metric and the electric field are expanded as a series of perturbations in the form:
\begin{align}
A(r) & = A_0 (r)+\xi A_1 (r)+\xi^2 A_2 (r)+\mathcal{O}(\xi^3), \label{Aap} \\
B(r) & = B_0 (r) + \xi B_1 (r)+\xi^2 B_2 (r)+\mathcal{O}(\xi^3), \label{Bap}  \\
E(r) & = E_0 (r) + \xi E_1 (r)+\xi^2 E_2 (r)+\mathcal{O}(\xi^3), \label{Eap}
\end{align}where we take as the background solution the BTZ (uncharged) metric and the Maxwell electric field, i.e.,  \begin{align}
A_0 (r) & = B_0 (r)= -M-\Lambda r^{2},\\
E_0 (r) & = \frac{Q}{r},
\label{A0E0}
\end{align}
and $\xi=\xi(a,b)$ is a small parameter used to guide the linearization of the equations of motion.

\subsection{First-Order Solution}

The field equations (\ref{Eq_gravitacional}), under the metric \textit{ansatz} (\ref{SS}) and the electromagnetic tensor (\ref{F}), have an intricate nonlinear form. However, by using the approximations (\ref{Aap})–(\ref{Eap}) up to first-order, we obtain the following linearized system of differential equations:
\begin{align}
& \frac{1}{2r}B_{1}'(r)+\frac{Q^{2}}{r^{2}(M+\Lambda r^{2})}\left(A_{1}(r)-B_{1}(r)\right)+\frac{2Q}{r}E_{1}(r)+\frac{Q^{2}}{\xi r^{4}}\big(r^{2}+4b^{2}(M-\Lambda r^{2})\big) =0, \label{Equations1}\\
& \frac{1}{2r}A_{1}'(r)+\frac{(Q^{2}-\Lambda r^{2})}{r^{2}(M+\Lambda r^{2})}\left(A_{1}(r)-B_{1}(r)\right)+\frac{2Q}{r}E_{1}(r)+\frac{Q^{2}}{\xi r^{4}}\left(r^{2}-4b^{2}(M+3\Lambda r^{2})\right)=0, \label{Equations2} \\
& \frac{1}{2}A_{1}''(r)-\frac{\Lambda r}{2(M+\Lambda r^{2})}\left(A_{1}'(r)-B_{1}'(r)\right)-\frac{\big(MQ^{2}+(M+Q^{2})\Lambda r^{2}\big)}{r^{2}(M+\Lambda r^{2})^{2}}\left(A_{1}(r)-B_{1}(r)\right)-\frac{2Q}{r}E_{1}(r) \nonumber \\& -\frac{Q^{2}}{\xi r^{4}}\left(r^{2}-4b^{2}(3M+\Lambda r^{2})\right)=0,  
\label{Equations3}
\end{align}where the prime $'$ denotes differentiation concerning $r$, and at this order we have neglected all terms involving factors such as $\xi a^2$ and $\xi b^2$ or $\xi^2$. Similarly to the four-dimensional case, all terms involving $a^2$ are of order $\xi^2$. Since $a^2$ or $b^2$ are assumed to be very small parameters, these terms are consequently disregarded \cite{Frizo:2022jyz}. Moreover, no higher-order derivative terms appear, and the electric field arises without derivatives. This simplification allows the system to be solved exactly, yielding the functions $A_1(r)$, $B_1(r)$, and $E_1(r)$ in the following form:
\begin{align}
&A_1(r)=c_{1}+\frac{c_{2}}{2}r^{2}+c_{3}\ln r-\frac{4\mathit{b}^{2}Q^{2}}{\xi r^{2}}\left(M+2\Lambda r^{2}+4\Lambda r^{2}\ln r\right), \label{A1}   \\
&B_1(r)=c_{1}-\frac{c_{2}M}{2\Lambda}+c_{3}\ln r+\frac{4\mathit{b}^{2}Q^{2}}{\xi r^{2}}\left(M-4\Lambda r^{2}\ln r\right), \label{B1}  \\
&E_1(r)=-\frac{Q c_2}{4\Lambda r}-\frac{c_3}{4Q r}+\frac{Q}{2\xi}\left(\frac{8\mathit{b}^{2}Q^{2}}{r^{3}}-\frac{1}{r}\left(1-12\mathit{b}^{2}\Lambda\right)\right), \label{E1}
\end{align}where $c_1$, $c_2$ and $c_3$ are integration constants. Since we are looking for small deviations from the charged BTZ geometry, namely,
\begin{equation}
    A(r)=B(r)=-M-\Lambda r^2-2Q^{2}\ln\left(\frac{r}{r_0}\right),
\end{equation}where $r_0$ is a positive scale parameter. So, we will set the integration constants as 
\begin{equation}
c_1=\frac{2Q^2}{\xi}\left(4 b^2 \Lambda +\ln r_0\right),\ \  c_2=0,\ \ \mbox{and}\ \  c_3=-\frac{2Q^2}{\xi}\left(1-8b^2\Lambda\right).\label{const1}    \end{equation}

Thus, the line element (\ref{SS}) and the electric field $E(r)$ can be written at first-order in our perturbative expansion as
 \begin{align}
 ds^2=& \Bigg[-M-\Lambda r^{2}-\frac{4b^{2}Q^{2}M}{r^{2}} -2Q^{2}\ln\left(\frac{r}{r_{0}}\right)\Bigg] dt^2 \nonumber \\
 -&\Bigg[-M-\Lambda r^{2}+8b^2Q^2\Lambda+\frac{4b^{2}Q^{2}M}{r^{2}} -2Q^{2}\ln\left(\frac{r}{r_{0}}\right)\Bigg]^{-1}dr^2 -r^2 d\theta^2,
 \label{Metric1}
 \end{align}and
 \begin{equation}
 E(r)=\frac{Q}{r}\left(1+2b^2\Lambda\right)+\frac{4b^2 Q^3}{r^3}.
 \label{Efinal1}
 \end{equation}
As is evident from the metric (\ref{Metric1}) and the electric field (\ref{Efinal1}), our choice of integration constants $c_i$ allows us to restore both the standard charged BTZ solution and the corresponding Maxwell electric field in the limit $b^2\to 0$. Furthermore, it can be observed that $A(r)\neq B(r)$, in contrast to the typical charged black hole solutions with spherical symmetry in general relativity. Such behavior is common in gravitational theories involving non-minimal vector couplings to curvature, as seen in models like bumblebee gravity \cite{Casana:2017jkc,Maluf:2020kgf}.

A glance at curvature invariants such as the Ricci and Kretschmann scalars ($R$ and $K$, respectively) calculated from the metric (\ref{Metric1}) yields
\begin{equation}
    R=-6\Lambda-\frac{2Q^2}{r^{2}}+\frac{16 b^{2} Q^{2}\Lambda}{r^{2}}-\frac{24 b^{2}Q^{2}M}{r^{4}},
\end{equation}
\begin{equation}
K=12\Lambda^{2}+\frac{8\Lambda Q^{2}}{r^{2}}-\frac{64b^2Q^2\Lambda^2}{r^2}+\frac{96\mathit{b}^{2}Q^{2}\Lambda M}{r^{4}},
\end{equation}which clearly differs from the charged BTZ invariants for nonnull $b^2$. We observe that the singularity at the origin $r=0$ remains unchanged, though the trend \footnote{Note that we say {\it trend} because we are using an expansion valid for large values of $r$, since otherwise the computed corrections would not be small.} of the Bopp-Podolsky corrections at this order is to increase its intensity from $1/r^2$ to $1/r^4$.  

\begin{figure}[!h]
\begin{center}
\begin{tabular}{ccc}
\includegraphics[height=7cm]{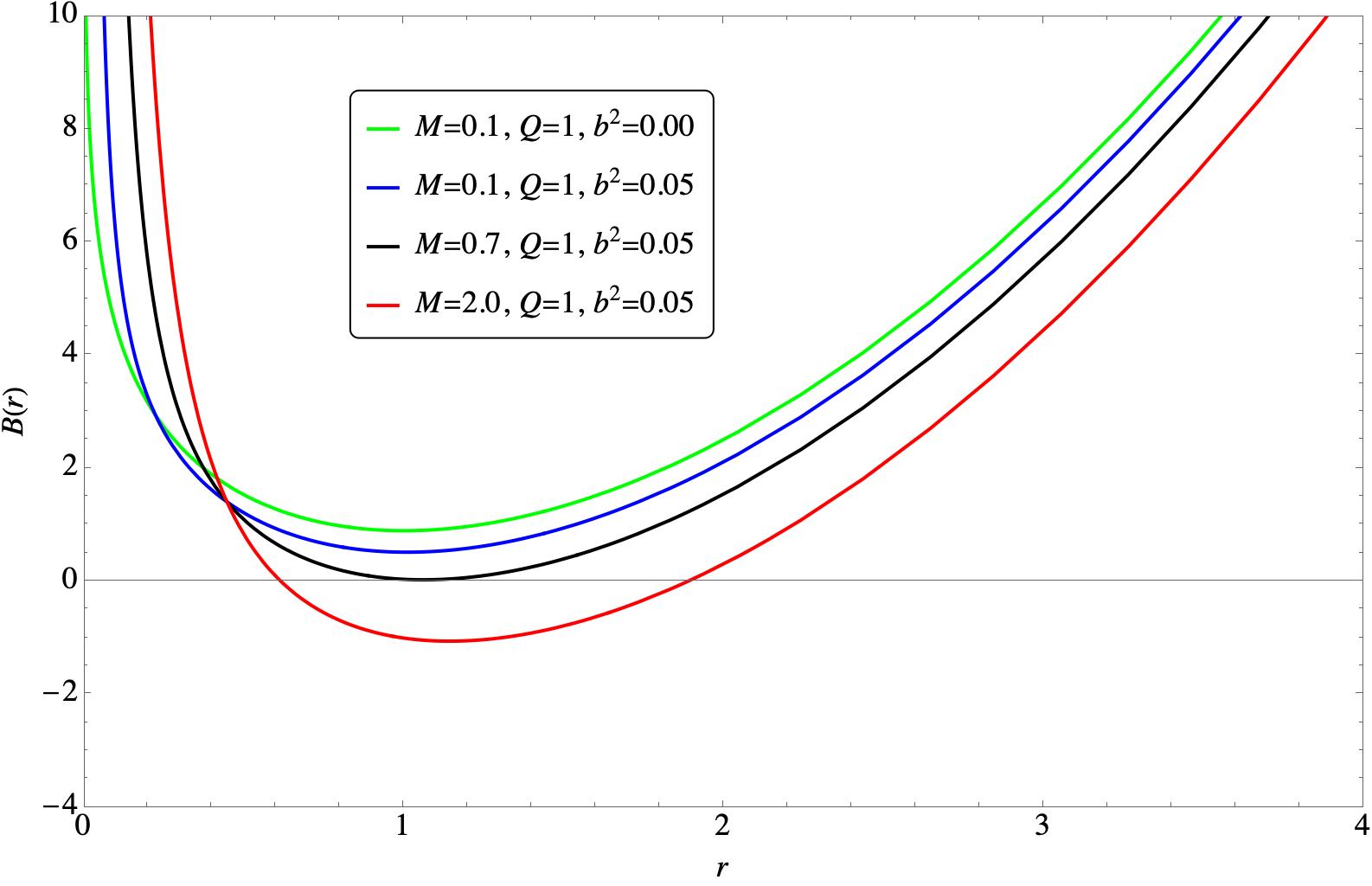}
\end{tabular}
\end{center}
\caption{Metric coefficient from Eq. (\ref{br}), as a function of the radial coordinate, for some values of $M$, $Q$, and $b^2$, with $\Lambda=-r^{-2}_{AdS}=-1$ and $r_0=1$.
\label{Figure1}}
\end{figure}

\begin{figure}[!h]
\begin{center}
\begin{tabular}{ccc}
\includegraphics[height=7cm]{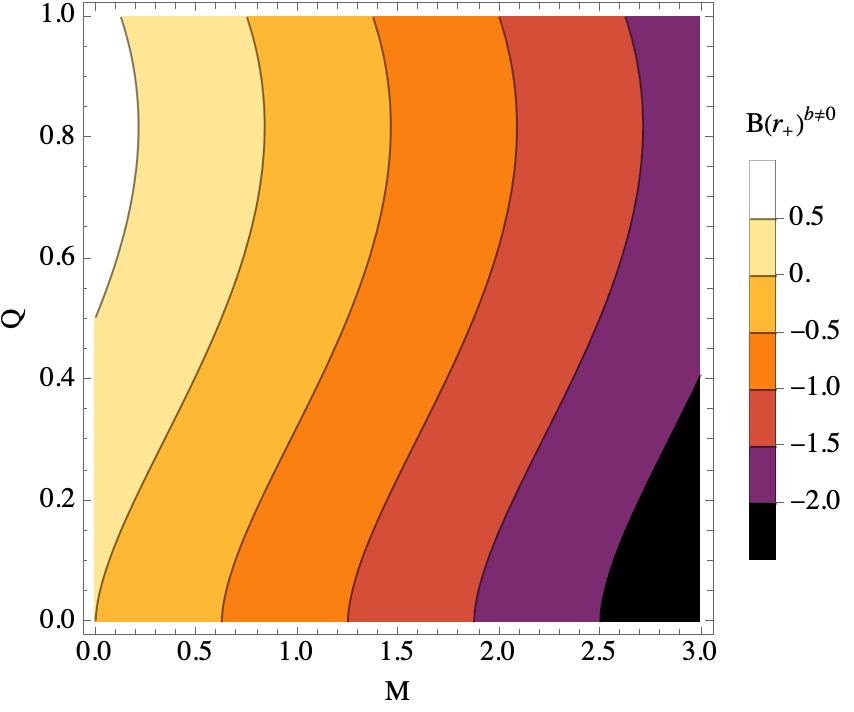}
\includegraphics[height=7cm]{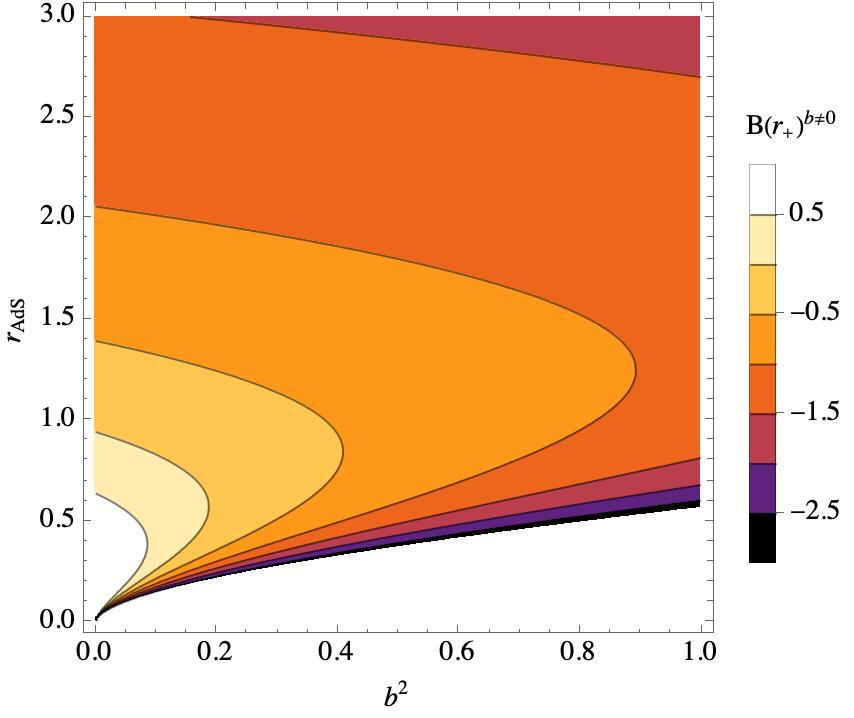}\\ 
(a) \hspace{6 cm}(b)\\
\end{tabular}
\end{center}
\caption{
Influence of the parameters $M,Q,b^2$, and $r_{AdS}$ on expression (\ref{Minimum1}). In these plots, we set $M = 0.1$ and $Q=0.8$ for the left panel, and $b^2 = 0.05$, $r_{AdS}=1$ for the right panel, with $r_0=1$. Regions with $B(r_+)^{b\neq 0}<0$ represent black hole solutions.\label{Figure2}}
\end{figure}

To analyze the spacetime causal structure, we focus on the localization of horizons. The positions of the horizons in spacetimes such as (\ref{Metric1}) are determined by the equation $g^{rr}=B(r)=0$. The function $B(r)$ can be expressed as
\begin{equation}
    B(r)=-M+8b^{2}Q^{2}\Lambda-\Lambda r^{2}-2Q^{2}\ln\left(\frac{r}{r_{0}}\right)+\frac{4b^{2}Q^{2}M}{r^2}\label{br} \ .
\end{equation}
Since in our approach the $b^2$ corrections must be small, this expression must be regarded as valid in the far region only (sufficiently large values of $r$), ceasing to be acceptable as $r\to 0$. The critical points of $B(r)$, where $dB/dr=0$, are located at
\begin{equation}\label{extrema}
r_{\pm}^2=\frac{{Q^2\pm\sqrt{Q^4+16Q^2 b^2 M }}}{{2(-\Lambda)}}\approx\frac{{Q^2\pm (Q^2+8 b^2 M )}}{{2(-\Lambda)}}  \ .
\end{equation}
When $b\to 0$, we get $r_{+}^2=Q^2/(-\Lambda)$ and $r_-=0$, which motivates the choice $\Lambda=-1/r^2_{AdS}$, with  $r_{AdS}$ representing the AdS radius. In that case, the function $B(r)$ only has an extremum (a minimum) at $r_+=r_{AdS} Q $, and diverges to $+\infty$ as $r\to \infty$ and as $r\to 0$ (due to the logarithm) \cite{Martinez:1999qi}. This qualitative structure can be preserved if the condition $M b^2>0$ is satisfied, because in that case the minimum still lies very close to $r_+$, being $r_-$ a complex number. If $M b^2<0$,  we may have a qualitatively different scenario, with $r_+\approx r_{AdS} Q  -2 |M b^2|/(r_{AdS} Q)$ and $r_-\approx 2\sqrt{|b^2M|}$. However, since the  new extremum at $r_-$ is expected to be close to zero because $|b^2M|\to 0$, it follows that  the perturbative expansion can not be trusted in that region and, therefore, such extremum must be regarded as an artifact of our approximation and should be discarded. The only relevant minima are thus given by $r_+\approx r_{AdS} Q  +2 M b^2/(r_{AdS} Q)$. Evaluating $B(r)$ at $r_{+}$, we get
\begin{equation}
B(r_+)^{b\neq 0}\equiv B(r_+)^{b= 0}+\frac{4b^2}{r_{AdS}^2}\left(M-2Q^2\right), \label{Minimum1}
\end{equation}
where 
\begin{equation}
 B(r_+)^{b= 0}=-M+Q^2-2Q^2\log\left(\frac{Q r_{AdS}}{r_0}\right)   
\end{equation}
Black holes exist whenever $B(r_+)^{b\neq 0}\leq 0$. If it is negative, there are two horizons (corresponding to two zeros of \( B(r) \)); if it is zero, the two roots coincide, resulting in an extremal black hole. These cases are depicted in Fig. \ref{Figure1}, where the standard scenario ($b^2$=0) is also illustrated.
The influence of the parameters $M, Q, b^2$, and $r_{AdS}$ on expression (\ref{Minimum1}) is illustrated in Figure \ref{Figure2}, where we consider $(b^2,r_{AdS}) =(0.05,1.0)$ (right panel) and  $(M,Q) =(1.0,0.8)$ (left panel). As one can see, for $0< M < 3$, there are regions where $B(r_+)^{b\neq 0}$ is positive, indicating the absence of black holes (white and yellowish bands on the upper left on the right panel). For the orange and darker bands, all solutions represent black holes. On the other hand, fixing $M$ and $Q$ (left panel), one sees that black hole solutions cease to exist in a region below $r_{AdS}<1$ and with $b^2$ smaller than $0.2$.

\subsubsection{Energy conditions}

Using the specific form of the electromagnetic field tensor (\ref{F}) with the electric field solution given by Eq. (\ref{Efinal1}), along with the metric (\ref{Metric1}), the linearized energy-momentum tensor $T_{\mu\nu}=T_{\mu\nu}^M+T_{\mu\nu}^a+T_{\mu\nu}^b$ in the Bopp–Podolsky electrodynamics can be expressed as a diagonal matrix, specifically:
\begin{equation}
    T^{\mu}_{\ \ \nu}=\begin{pmatrix}
    \rho(r)\\
 & -p_{r}(r)\\
 &  & -p_{\theta}(r)
\end{pmatrix},
\end{equation}
 with
\begin{equation}
    \rho(r)=\frac{Q^2}{8\pi r^2}\left(1+\frac{4b^2 M}{r^2}\right)\geq0,
\end{equation}

\begin{equation}
    p_r(r)=\frac{Q^2}{8\pi r^2}\left(-1+8b^2 \Lambda+\frac{4b^2 M}{r^2}\right),\label{pr1}
\end{equation}

\begin{equation}
    p_\theta(r)=\frac{Q^2}{8\pi r^2}\left(1-\frac{12b^2 M}{r^2}\right),\label{ptheta1}
\end{equation}
where $\rho$ and $p_r,p_\theta$ are the density and effective pressures, respectively, of the Bopp–Podolsky field. As we can see, the Bopp–Podolsky energy-momentum tensor exhibits its anisotropic feature as evidenced by $p_r\neq p_{\theta}$. We also note that, unlike in nonlinear theories of electrodynamics, the condition $p_r=-\rho$ is not satisfied when $b^2$ corrections are considered. This property also justifies the fact that $A(r)\neq B(r)$. 

 Specifically, the weak energy condition (WEC) requires $\rho(r) > 0$ and $\rho + p_i \geq 0$ (for $i = r$ or $\theta$). The null energy condition (NEC) stipulates that the sum of the energy density and any pressure satisfies $\rho + p_i \geq 0$, while the dominant energy condition (DEC) states that the energy density must be greater than or equal to the absolute value of any of the pressures, $\rho - |p_i| \geq 0$. From the expressions provided above, we find that 
\begin{eqnarray}
    \rho+p_r&=& \frac{Q^2b^2}{\pi r^2}\left(\frac{M}{r^2}-\frac{1}{r_{AdS}^2}\right) \\
    \rho+p_\theta&=& \frac{Q^2}{4\pi r^2}\left(1-\frac{4b^2M}{r^2}\right) \\
     \rho-|p_r|&=& \frac{Q^2}{4\pi r^2}\left(1+\frac{4b^2}{r_{AdS}^2}\right) \\
    \rho-|p_\theta|&=& \frac{2Q^2b^2M}{\pi r^4} \\
    \rho+p_r+p_\theta&=& \frac{Q^2}{8\pi r^2}\left(1-\frac{8b^2}{r_{AdS}^2}-\frac{4b^2M}{r^2}\right) \ ,
\end{eqnarray}
which indicate that the energy conditions are not universally satisfied across the domain in the event of having $b^2M<0$, with violations of WEC, NEC, and DEC. Only when $b^2M\geq 0$ are the energy conditions satisfied. Additionally, the strong energy condition (SEC), which necessitates that NEC holds and $\rho + \sum_i p_{i} \geq 0$, is also violated due to the breakdown of NEC.

We have just seen that, like in the four-dimensional context considered in Refs. \cite{Cuzinatto:2017srn,Frizo:2022jyz}, the Bopp-Podolsky electrodynamics is able to violate the energy conditions also in $2+1$ dimensions. The main implication of this result is that violations of energy conditions could lead to tachyons or signals propagating faster than light in a vacuum \cite{Shabad:2011hf}. The energy conditions also provide information about the cause for the existence of wormholes. However, since we are dealing with a perturbative expansion, one should be cautious when extrapolating our results, valid in the far region, to the innermost region of the space-time, where higher-order corrections would be necessary. A more in-depth analysis of the properties of photons and their trajectories in this theory will be carried out elsewhere, as it lies beyond the scope of this paper. 


\subsection{Second Order Solution}

In order to establish on more solid grounds the trend of the metric components in the far region, it is useful to compute the next-order corrections. The second-order perturbations $A_2$, $B_2$, and $E_2$ are governed by linear equations that account for the interactions between the first-order perturbations. These perturbations were expressed in Eqs.(\ref{Aap}),(\ref{Bap}) and (\ref{Eap})
where $A_1$, $B_1$, and $E_1$ are determined by Eqs. (\ref{A1})–(\ref{E1}), with the integration constants fixed as specified in Eq. (\ref{const1}).

Substituting the above expressions into the field equations under the metric \textit{ansatz} (\ref{SS}), linearizing around $\xi = 0$, and collecting terms up to order $\xi^2$, we derive the following set of linear differential equations:
\begin{align}
& \frac{1}{2r}B_{2}'(r)+\frac{Q^{2}}{r^{2}(M+\Lambda r^{2})}\left(A_{2}(r)-B_{2}(r)\right)+\frac{2Q}{r}E_{2}(r) =f(r), \label{Equations12}\\
& \frac{1}{2r}A_{2}'(r)+\frac{(Q^{2}-\Lambda r^{2})}{r^{2}(M+\Lambda r^{2})}\left(A_{2}(r)-B_{2}(r)\right)+\frac{2Q}{r}E_{2}(r)=g(r), \label{Equations22} \\
& \frac{1}{2}A_{2}''(r)-\frac{\Lambda r}{2(M+\Lambda r^{2})}\left(A_{2}'(r)-B_{2}'(r)\right)-\frac{\big(MQ^{2}+(M+Q^{2})\Lambda r^{2}\big)}{r^{2}(M+\Lambda r^{2})^{2}}\left(A_{2}(r)-B_{2}(r)\right)-\frac{2Q}{r}E_{2}(r) \nonumber \\& =h(r),  
\label{Equations32}
\end{align}where

\begin{align}
f(r)&=\frac{4\mathit{b}^{2}Q^{2}}{\xi^{2}r^{6}\left(M+\Lambda r^{2}\right)}\left[2MQ^{2}r^{2}+4\mathit{b}^{2}\left(Q^{2}+M\right)\left(Q^{2}M-\Lambda r^{2}\left(M-3Q^{2}\right)\right)\right.\nonumber\\&\left.+\Lambda r^{4}\left(2Q^{2}+\mathit{b}^{2}\Lambda\left(-M+4Q^{2}+3\Lambda r^{2}\right)\right)-2Q^{2}r^{2}\left(M+2Q^{2}+\Lambda r^{2}\right)\ln\left(\frac{r}{r_{0}}\right)\right],
\end{align}

\begin{align}
g(r)&=\frac{4\mathit{b}^{2}Q^{2}}{\xi^{2}r^{6}\left(M+\Lambda r^{2}\right)}\left[4\mathit{b}^{2}\left(\Lambda r^{2}\left(M^{2}+3Q^{4}\right)+MQ^{2}\left(M+Q^{2}\right)\right)\right.\nonumber\\&\left.+\Lambda r^{4}\left(\mathit{b}^{2}\Lambda\left(15M-12Q^{2}+11\Lambda r^{2}\right)+6Q^{2}\ln\left(\frac{r}{r_{0}}\right)\right)-2Q^{2}r^{2}\left(2Q^{2}-M\right)\ln\left(\frac{r}{r_{0}}\right)\right],
\end{align}

\begin{align}
    h(r)&=-\frac{4\mathit{b}^{2}Q^{2}}{\xi^{2}r^{6}\left(M+\Lambda r^{2}\right)}\left\{2M\left[2\mathit{b}^{2}\left(\Lambda r^{2}\left(3M^{2}+9MQ^{2}+4Q^{4}\right)+MQ^{2}\left(5M+Q^{2}\right)\right)-3MQ^{2}r^{2}\right]\right.\nonumber\\&+\Lambda r^{4}\left(3\mathit{b}^{2}\Lambda\left(9M^{2}+M\left(4Q^{2}+6\Lambda r^{2}\right)+\left(\Lambda r^{2}-2Q^{2}\right)^{2}\right)-2Q^{2}\left(7M+4\Lambda r^{2}\right)\right)\nonumber\\&\left.+2Q^{2}r^{2}\left(3M^{2}-2M\left(Q^{2}-3\Lambda r^{2}\right)-\Lambda r^{2}\left(2Q^{2}-5\Lambda r^{2}\right)\right)\ln\left(\frac{r}{r_{0}}\right)\right\}.
\end{align}

As before, we have suppressed all terms involving factors such as $\xi^2a^2$, $\xi^2b^2$, or $\xi^3$. It is noteworthy that the left-hand side of the above equations retains the same structure as the first-order equations (\ref{Equations1})–(\ref{Equations3}), while the right-hand side differs by incorporating corrections arising from the first-order fields. Once again, we discard terms like $a^2\xi^2$, $b^2\xi^2$, and $\xi^3$ or higher-order contributions. Furthermore, similar to the previous case, there are no contributions from the coefficient $a^2$, as it always appears in higher-order terms.

As in the previous case, the linear system presented in Eqs. (\ref{Equations12})–(\ref{Equations32}) is exactly solvable, and its solutions are given by:
\begin{align}
&A_2(r)=c_{4}+\frac{c_{5}}{2}r^{2}+c_{6}\ln r\nonumber\\
&-\frac{1}{\xi^2 }\left[\frac{16b^4 Q^2\Lambda M}{r^2}+32b^4 Q^2\Lambda ^2 \left(1+2\ln r\right)-\frac{4b^2 Q^4}{r^4}\left(r^2-2b^2M-2r^2\ln\left(\frac{r}{r_0}\right)\right)\right], \label{A2}   \\
&B_2(r)=c_{4}-\frac{c_{5}M}{2\Lambda}+c_{6}\ln r\nonumber\\
&+\frac{1}{\xi^2 }\left[\frac{16b^4 Q^2\Lambda (M-2Q^2)}{r^2}-64 b^4 Q^2\Lambda ^2 \ln r +\frac{4b^2 Q^4}{r^4}\left(r^2-2b^2M+2r^2\ln\left(\frac{r}{r_0}\right)\right)\right], \label{B2}   \\
& E_2(r)=-\frac{Q c_5}{4\Lambda r}-\frac{c_6}{4Q r}+\frac{1}{\xi ^2}\left[ \frac{22b^4\Lambda^2 Q}{r}+\frac{4 b^2 Q^3 \left(1+2 b^2\Lambda\right)}{r^3}+\frac{8b^4 Q^5}{r^5}\right].
\end{align} 

As expected, the structure of the solutions to the homogeneous part has the same form as in the first-order case, with $c_4$, $c_5$, and $c_6$ being integration constants. To preserve the structure of the first-order solution, we can choose the integration constants as follows:
\begin{equation}
    c_{4}=\frac{32b^{4}Q^{2}\Lambda^{2}}{\xi^{2}},\ \ c_{5}=0,\ \ c_{6}=\frac{64b^{4}Q^{2}\Lambda^{2}}{\xi^{2}}
\end{equation}
We can now write the line element and the electric field for the Einstein-Bopp-Podolsky system, incorporating corrections up to second-order perturbations, as:
\begin{align}
    ds^{2}	&=\left[-M-\Lambda r^{2}-\frac{4b^{2}Q^{2}}{r^{2}}\left(M+4b^{2}M\Lambda-Q^{2}\right)-\frac{8b^{4}Q^{4}M}{r^{4}}-\left(2Q^{2}+\frac{8b^{2}Q^{4}}{r^{2}}\right)\ln\left(\frac{r}{r_{0}}\right)\right]dt^{2}\nonumber\\
	&-\left[-M-\Lambda r^{2}+8b^{2}Q^{2}\Lambda\left(1+4b^{2}\Lambda\right)+\frac{4b^{2}Q^{2}}{r^{2}}\left(M+Q^{2}+4b^{2}\Lambda (M-2Q^2)\right)\right.\nonumber\\
	&\left. -\frac{8b^{4}Q^{4}M}{r^{4}}-\left(2Q^{2}-\frac{8b^{2}Q^{4}}{r^{2}}\right)\ln\left(\frac{r}{r_{0}}\right)\right]^{-1}dr^{2}
	-r^{2}d\theta^{2},
\end{align}
\begin{equation}
    E(r)=\frac{Q}{r}\left(1+2b^{2}\Lambda+6b^{4}\Lambda^{2}\right)+\frac{8b^{2}Q^{3}}{r^{3}}\left(1+b^{2}\Lambda\right)+\frac{8b^{4}Q^{5}}{r^{5}}.
\end{equation}
From this expression we see that all the corrections induced by the $b$ parameter decay very fast in the far region and suggest that the BTZ zeroth-order solution is modified by series expansions with negative powers. Note that this also happens in the logarithmic term.

\section{Conclusion}\label{conclusion}

 In this work we have explored corrections induced to the circularly symmetric BTZ solution of GR when the electric field is governed by Bopp-Podolsky electrodynamics, a gauge invariant theory with higher-order derivatives. The complexity of the resulting set of field equations motivated a perturbative analysis and, by computing the first-order corrections in the Bopp-Podolsky coupling parameters, we have shown that black hole configurations are still possible, though the structure of horizons is slightly modified with respect to the BTZ solution. The energy conditions have also been investigated, providing evidence for violations when $b^2M<0$ in regions where the perturbative expansions are reliable. Nonetheless, the amplitude of such violations is small, and one cannot claim that they can support strong modifications of the geometry, like turning black hole configurations into wormholes \cite{Frizo:2022jyz}. In fact, we have seen that first-order and second-order corrections are small compared to the background BTZ solution in the far region and that they point towards a decaying trend in the higher-order contributions of the perturbative expansion. Though the functional form of the computed corrections suggests an important growth in the $r\to 0$ region, one cannot use them to extract any quantitative conclusions about changes in the space-time in that limit because higher-order corrections would become dominant and necessary in the analysis. For such purpose, a non-perturbative numerical integration of the field equations would be necessary. We hope to address this aspect in a follow up of this work.


\section*{Acknowledgments}
\hspace{0.5cm} The authors thank the Funda\c{c}\~{a}o Cearense de Apoio ao Desenvolvimento Cient\'{i}fico e Tecnol\'{o}gico (FUNCAP), and the Conselho Nacional de Desenvolvimento Cient\'{i}fico e Tecnol\'{o}gico (CNPq), Grant no. 200879/2022-7 (RVM) for financial support. 
The authors also acknowledge financial support from the Spanish Grants  PID2020-116567GB-C21, PID2023-149560NB-C21 funded by MCIN/AEI
/10.13039/501100011033, and by CEX2023-001292-S funded by MCIU/AEI.  The paper is based upon work from COST Action CaLISTA CA21109 supported by COST (European Cooperation in Science and Technology).

\vspace{0.5cm}
\center{\bf No data associated in the manuscript}

\appendix



\begin{thebibliography}{99}



\bibitem{Bopp1940}
  Fritz Bopp,
  Annalen der Physik, {\bf 430} (1940) 345-384
  doi.org/10.1002/andp.19404300504
  

\bibitem{Podolsky:1942zz}
B.~Podolsky,
Phys. Rev. \textbf{62} (1942), 68-71
doi:10.1103/PhysRev.62.68

\bibitem{Podolsky:1944zz}
B.~Podolsky and C.~Kikuchi,
Phys. Rev. \textbf{65} (1944), 228-235
doi:10.1103/PhysRev.65.228


\bibitem{Proca}
A. Proca, J. Phys. Radium \textbf{7}, 347 (1936).


\bibitem{Tu:2005ge}
L.~C.~Tu, J.~Luo and G.~T.~Gillies,
Rept. Prog. Phys. \textbf{68} (2005), 77-130
doi:10.1088/0034-4885/68/1/R02


\bibitem{Accioly:2010zza}
A.~Accioly and E.~Scatena,
Mod. Phys. Lett. A \textbf{25} (2010), 269-276
doi:10.1142/S0217732310031610

\bibitem{Kaparulin:2014vpa}
D.~S.~Kaparulin, S.~L.~Lyakhovich and A.~A.~Sharapov,
Eur. Phys. J. C \textbf{74} (2014) no.10, 3072
doi:10.1140/epjc/s10052-014-3072-3
[arXiv:1407.8481 [hep-th]].

\bibitem{Galvao:1986yq}
C.~A.~P.~Galvao and B.~M.~Pimentel Escobar,
Can. J. Phys. \textbf{66} (1988), 460-466
doi:10.1139/p88-075

\bibitem{Ferreira:2019lpu}
M.~M.~Ferreira, L.~Lisboa-Santos, R.~V.~Maluf and M.~Schreck,
Phys. Rev. D \textbf{100} (2019) no.5, 055036
doi:10.1103/PhysRevD.100.055036
[arXiv:1903.12507 [hep-th]].


\bibitem{Borges:2019gpz}
L.~H.~C.~Borges, F.~A.~Barone, C.~A.~M.~de Melo and F.~E.~Barone,
Nucl. Phys. B \textbf{944} (2019), 114634
doi:10.1016/j.nuclphysb.2019.114634
[arXiv:1906.02741 [hep-th]].

\bibitem{Bufalo:2012tt}
R.~Bufalo, B.~M.~Pimentel and G.~E.~R.~Zambrano,
Phys. Rev. D \textbf{86} (2012), 125023
doi:10.1103/PhysRevD.86.125023
[arXiv:1212.3542 [hep-th]].

\bibitem{Bufalo:2010sb}
R.~Bufalo, B.~M.~Pimentel and G.~E.~R.~Zambrano,
Phys. Rev. D \textbf{83} (2011), 045007
doi:10.1103/PhysRevD.83.045007
[arXiv:1008.3181 [hep-th]].

\bibitem{Bonin:2009je}
C.~A.~Bonin, R.~Bufalo, B.~M.~Pimentel and G.~E.~R.~Zambrano,
Phys. Rev. D \textbf{81} (2010), 025003
doi:10.1103/PhysRevD.81.025003
[arXiv:0912.2063 [hep-th]].

\bibitem{AraujoFilho:2020bzd}
A.~A.~Ara\'ujo Filho and R.~V.~Maluf,
Braz. J. Phys. \textbf{51} (2021) no.3, 820-830
doi:10.1007/s13538-021-00880-0
[arXiv:2003.02380 [hep-th]].

\bibitem{Bonin:2016gav}
C.~A.~Bonin, B.~M.~Pimentel and P.~H.~Ortega,
Int. J. Mod. Phys. A \textbf{34} (2019) no.24, 1950134
doi:10.1142/S0217751X19501343
[arXiv:1608.00902 [hep-th]].

\bibitem{Cuzinatto:2017srn}
R.~R.~Cuzinatto, C.~A.~M.~de Melo, L.~G.~Medeiros, B.~M.~Pimentel and P.~J.~Pompeia,
Eur. Phys. J. C \textbf{78} (2018) no.1, 43
doi:10.1140/epjc/s10052-018-5525-6
[arXiv:1706.09455 [gr-qc]].


\bibitem{Frizo:2022jyz}
D.~A.~Frizo, C.~A.~M.~de Melo, L.~G.~Medeiros and J.~C.~S.~Neves,
Annals Phys. \textbf{457} (2023), 169411
doi:10.1016/j.aop.2023.169411
[arXiv:2210.09938 [gr-qc]].

\bibitem{Cuzinatto:2016kjk}
R.~R.~Cuzinatto, E.~M.~de Morais, L.~G.~Medeiros, C.~Naldoni de Souza and B.~M.~Pimentel,
EPL \textbf{118} (2017) no.1, 19001
doi:10.1209/0295-5075/118/19001
[arXiv:1611.00877 [astro-ph.CO]].

\bibitem{Kruglov:2009yr}
S.~I.~Kruglov,
J. Phys. A \textbf{43} (2010), 245403
doi:10.1088/1751-8113/43/24/245403
[arXiv:0907.1706 [hep-th]].

\bibitem{Cuzinatto:2011zz}
R.~R.~Cuzinatto, C.~A.~M.~de Melo, L.~G.~Medeiros and P.~J.~Pompeia,
Int. J. Mod. Phys. A \textbf{26} (2011), 3641-3651
doi:10.1142/S0217751X11053961
[arXiv:0810.4106 [quant-ph]].

\bibitem{Zayats:2013ioa}
A.~E.~Zayats,
Annals Phys. \textbf{342} (2014), 11-20
doi:10.1016/j.aop.2013.12.005
[arXiv:1306.3966 [hep-th]].

\bibitem{Granado:2019bqk}
D.~R.~Granado, A.~J.~G.~Carvalho, A.~Y.~Petrov and P.~J.~Porfirio,
EPL \textbf{129} (2020) no.5, 51001
doi:10.1209/0295-5075/129/51001
[arXiv:1912.00855 [hep-th]].

\bibitem{Israel:1967wq}
W.~Israel,
Phys. Rev. \textbf{164} (1967), 1776-1779
doi:10.1103/PhysRev.164.1776

\bibitem{Israel:1967za}
W.~Israel,
Commun. Math. Phys. \textbf{8} (1968), 245-260
doi:10.1007/BF01645859

\bibitem{Carter:1971zc}
B.~Carter,
Phys. Rev. Lett. \textbf{26} (1971), 331-333
doi:10.1103/PhysRevLett.26.331

\bibitem{Casana:2017jkc}
R.~Casana, A.~Cavalcante, F.~P.~Poulis and E.~B.~Santos,
Phys. Rev. D \textbf{97} (2018) no.10, 104001
doi:10.1103/PhysRevD.97.104001
[arXiv:1711.02273 [gr-qc]].

\bibitem{Maluf:2020kgf}
R.~V.~Maluf and J.~C.~S.~Neves,
Phys. Rev. D \textbf{103} (2021) no.4, 044002
doi:10.1103/PhysRevD.103.044002
[arXiv:2011.12841 [gr-qc]].


\bibitem{Martinez:1999qi}
C.~Martinez, C.~Teitelboim and J.~Zanelli,
Phys. Rev. D \textbf{61} (2000), 104013
doi:10.1103/PhysRevD.61.104013
[arXiv:hep-th/9912259 [hep-th]].

\bibitem{Shabad:2011hf}
A.~E.~Shabad and V.~V.~Usov,
Phys. Rev. D \textbf{83} (2011), 105006
doi:10.1103/PhysRevD.83.105006
[arXiv:1101.2343 [hep-th]].














\end{thebibliography}
\end{document}